\documentclass[a4paper]{article}

\usepackage{icrc2013}
\usepackage[english]{babel}

\title{Discovery of the mysterious gamma-ray source HESSJ1832-093 in the vicinity of SNR G22.7-0.2}

\shorttitle{Discovery of the gamma-ray source HESSJ1832-093}

\authors{
H. Laffon$^{1,2}$,
F. Acero$^{3}$,
F. Brun$^{4}$,
B. Kh\'elifi$^{2}$,
G. P\"uhlhofer$^{5}$,
R. Terrier$^{6}$,
for the H.E.S.S. Collaboration.
}

\afiliations{
\scriptsize{
$^1$ Centre d'Etudes Nucl\'eaires de Bordeaux-Gradignan, CNRS/IN2P3, Universit\'e Bordeaux 1, France \\
$^2$ Laboratoire Leprince-Ringuet, Ecole Polytechnique, CNRS/IN2P3, Palaiseau, France \\
$^3$ NASA Goddard Space Flight Center, Greenbelt, USA\\
$^4$ Max-Planck-Institut f\"ur Kernphysik, Heidelberg, Germany \\
$^5$ Institut f\"ur Astronomie und Astrophysik, Universit\"at T\"ubingen, Germany \\
$^6$ Astroparticules et Cosmologie, CNRS/IN2P3, Universit\'e Paris 7, Paris, France\\
}
}

\email{laffon@cenbg.in2p3.fr}

\abstract{
Thanks to the use of advanced analysis techniques, the H.E.S.S. imaging Cherenkov telescope array has now reached a sensitivity level allowing the detection of sources with fluxes around 1$\%$ of the Crab with a limited observation time (less than $\sim$100 hours). 67 hours of observations in the region of the supernova remnant G22.7-0.2 thus yielded the detection of the faint point-like source HESS J1832-093 in spatial coincidence with a part of the radio shell of the supernova remnant. A multi-wavelength search for counterparts was then performed and led to the elaboration of various scenarii in order to explain the origin of the very high energy excess. 
The presence of molecular clouds in the line of sight could indicate a hadronic origin through $\pi^0$ production and decay. However, the discovery of an X-ray point source and its potential infrared counterpart give rise to other possibilities such as a pulsar wind nebula nature or a binary system. The latest results on this source are presented as well as the different scenarii brought by the multi-wavelength observations.
}


\begin{document}

\maketitle

\section{Introduction}

H.E.S.S. (High Energy Stereoscopic System) is an array of four imaging atmospheric Cherenkov telescopes located 
$1800~ \rm m$ above sea level in the Khomas Highland of Namibia and has been fully operational since 2004 \cite{crab}.
The H.E.S.S. collaboration has been conducting a systematic scan of the Galactic Plane, which led to the discovery of a 
rich population of very high energy (VHE) gamma-ray sources such as in
the region around the supernova remnant (SNR) W41.

As a consequence, a dedicated observation campaign was launched by H.E.S.S. to study those TeV sources in detail.
These observations led to the discovery of HESS J1832$-$093 in spatial coincidence with the edge of SNR G22.7$-$0.2. 
The latter shows a non-thermal shell of 26$'$ diameter in radio \cite{Shaver} and partially overlaps the neighbouring remnant W41.
A search for multiwavelength counterparts was performed in order to identify the nature of HESS J1832$-$093. 
A proposal for observations in X-rays at the position of HESS J1832$-$093 was submitted to the XMM-\textit{Newton} satellite and 
allowed the discovery of a point-like source in X-rays near the center of the TeV source. 
Apart from the radio emission of the SNR, no radio point-like counterpart was found but measurements of the $^{13}$CO (J=1$\rightarrow$0) 
transition line show the presence of molecular structures on the line of sight. 
However, an infrared point-like source lying less than 2$''$ away from the center of the X-ray source was found in the 2MASS catalog.
The features of the TeV emission discovered by H.E.S.S. are described in section 2. 
Details on the X-ray, IR and radio counterparts are given in section 3 and different considered 
scenarios to explain the VHE gamma-ray emission are presented in section 4.

\section{H.E.S.S. observations}

\subsection{Detection and morphology}
A standard run selection procedure is used to remove bad quality observations in order to study the field of view. 
This results in a data set comprising 67 hours live time of observations taken from 2004 to 2011.

\begin{figure}[!htbp]
  \centering
\includegraphics[width=0.45\textwidth,bb=95 195 502 598,clip]{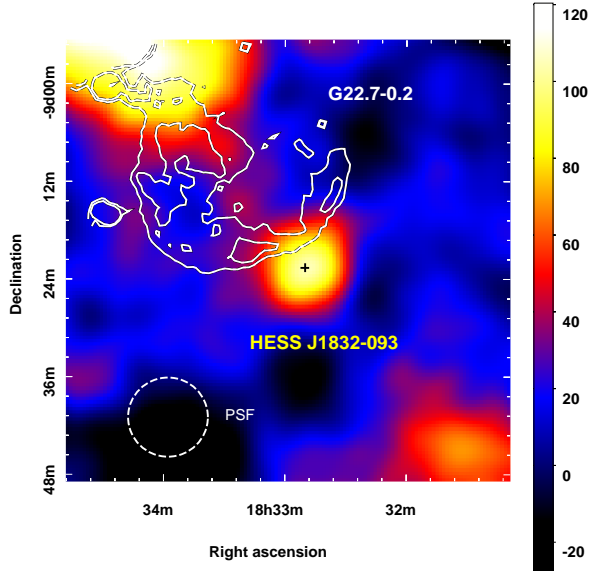}
  \caption{H.E.S.S. excess map oversampled with the r$_{68}$ value of 0.081$^\circ$ (dashed circle) and smoothed with a 2D-gaussian 
with the same width in units of counts per integration area. The newly discovered source HESS J1832$-$093 is shown with its best-fit position 
and corresponding statistical errors (black cross). The SNR  G22.7$-$0.2 observed in radio  \cite{magpis} is represented by the white contours. 
The emission seen on the upper left is a small part of HESS J1834$-$087 \cite{hess2006survey}, the TeV source in spatial coincidence with SNR W41. 
}
  \label{HESSexcess}
 \end{figure}

A standard H.E.S.S. analysis frame is adopted with the Hillas reconstruction method using the weighted intersections of the  
main axis of the images to reconstruct the direction of the events \cite{crab}. A recently developped multi-variate analysis is used to 
provide a better discrimination between hadrons and gamma-rays \cite{parismva}. A minimum charge of 110 photo-electrons in the shower images 
is applied to the data, resulting in an energy threshold of about 450 GeV. 
The background is estimated with the ring-background model, as described in \cite{crab}. 

Using these techniques, HESS J1832$-$093 is detected with a peak significance of $7.9~\sigma$ pre-trials, resulting in a post-trial 
detection significance of $5.6~\sigma$, following the approach described in \cite{hess2006survey}.
The corresponding excess map of the field of view showing the new detected source is presented in Fig. \ref{HESSexcess}.
The average angular resolution (r$_{68}$) for this selected data set and assuming a power-law spectral index of 
2.3 is $0.081^\circ$ at the source position.
A two-dimensional symmetrical Gaussian function is used to determine the position and size of the TeV emission with 
a $\chi^2$ minimization. 
The best-fit position obtained is $\rm RA=18^h 32^m 50^s \pm 32^s_{stat} \pm 36^s_{syst}, \rm  Dec=-9^\circ 22' 36'' \pm 32''_{stat} \pm 36''_{syst}~(J2000)$ ($\chi^2$/ndf=0.89).
No significant extension was found for the source.

\subsection{Spectral analysis}

To produce the energy spectrum, only the highest quality data are used, corresponding to a data set of $59$ hours live time.
In order to broaden the accessible energy range,
the charge cut is lowered to a minimum of 80 photo-electrons, resulting in an energy threshold of $\sim$400 GeV.
The background is estimated with the reflected background 
model, which is better suited for spectral studies \cite{crab}. 
The forward-folding method described in \cite{crab} is applied to the data to derive the spectrum.
Source counts are extracted from a circular region of 0.1$^\circ$ radius around the best fit position of HESS J1832$-$093, a size optimized for 
point source studies with the applied cuts.
The corresponding excess is $152 \pm 24$ gamma-like events.

\begin{figure}[!htbp]
  \centering
  \includegraphics[width=0.45\textwidth]{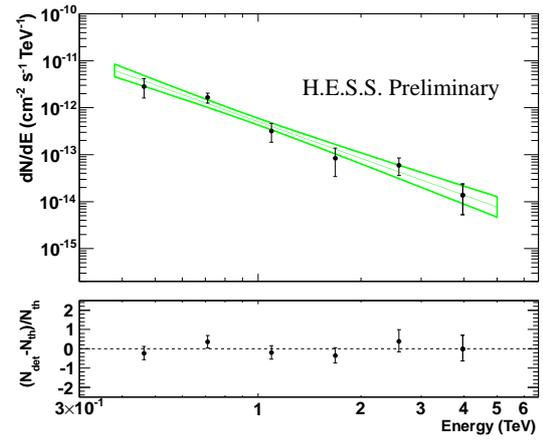}
  \caption{Top panel: VHE gamma-ray spectrum observed from HESS J1832$-$093 obtained with the forward-folding method. Bottom panel: corresponding residuals. 
The green contour represents the 1$\sigma$ confidence level of the fitted spectrum using a power-law hypothesis. 
}
  \label{spectrum}
 \end{figure}

The spectrum obtained between 400 GeV and 5 TeV (displayed on Fig. \ref{spectrum}) can be described by a power-law 
${{d\Phi} \over {dE}}\,=\,\Phi_0 \left( {{E} \over {1\,\rm TeV}} \right) ^{-\Gamma}$,
with an index $\Gamma = 2.6 \pm 0.3_{\rm stat} \pm 0.3_{\rm syst}$ and a differential flux normalisation at 1 TeV of 
$\Phi_0=(4.8 \pm 0.8_{\rm stat}\pm 0.9_{\rm syst})\,\times\,10^{-13}\,\rm{cm} ^{-2}\,s^{-1}\,TeV^{-1}$.
A light curve was produced with the available observations and no significant temporal variability was detected in the H.E.S.S. data set.

\section{Multi-wavelength observations}

\subsection{X-Ray observations with XMM-\textit{Newton}}\label{xmm}

In order to constrain the nature of the source HESS J1832$-$093, a dedicated XMM-\textit{Newton} observation (ID: 0654480101) was performed 
in March 2011 for 17 ks, targeted at the position of the gamma-ray emission.
After filtering out proton flare contamination, 13 ks  and 7 ks of exposure time remained for the two EPIC-MOS cameras and for the EPIC-pn camera respectively.

The data were processed using the XMM-\textit{Newton} Science Analysis System (v10.0). The instrumental background
 was derived from a compilation of blank sky observations \cite{carter07}, renormalized to the actual exposure using the count rate in
the 10-12 keV energy band.

\begin{figure}[!htbp]
  \centering
 \includegraphics[width=0.45\textwidth,bb=58 206 484 586,clip]{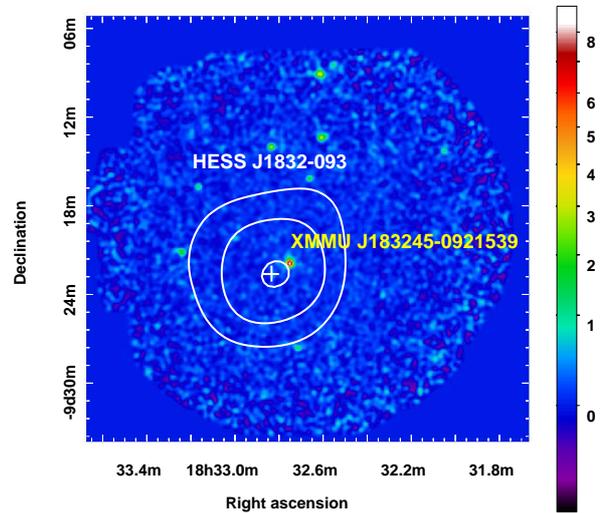}
  \caption{X-ray count map in the 2-10 keV band smoothed with a Gaussian of 15$''$ width (HPD of the EPIC cameras PSF) of the field of view 
around HESS J1832$-$093 as seen by XMM-\textit{Newton}. The white cross symbolizes the best-fit position of HESS J1832$-$093 with corresponding 
statistical errors and the white contours represent the H.E.S.S. excess.}
  \label{xraymap}
 \end{figure}

The brightest object in the XMM-\textit{Newton} field of view is a point-like source located at 
$\rm RA=18^h 32^m 45^s.04, \rm Dec=-09^\circ 21' 53''.9 $, 
1.4$'$ away from the best-fit position of the H.E.S.S. excess, as  shown in Fig. \ref{xraymap}. 
This new source, dubbed XMMU J183245$-$0921539, lies within the 2$\sigma$ uncertainty of the VHE gamma-ray centroid.

Spectra from the three instruments were extracted from a 15$''$ radius circular 
region centered on XMMU J183245$-$0921539. 
As the statistics are low ($\sim$500 counts when summing the three EPIC instruments), 
the data were fitted in Xspec (v12.5) using Cash statistics and the spectra were not rebinned for the fitting process. 
The p-value corresponding to the null-hypothesis probability is derived using the \textit{goodness} command  in XSPEC which performs Monte Carlo  
simulations of the goodness-of-fit.
The best fit parameters for a power-law assumption are a column density
N${_{\rm H}}= 10.5^{+3.1}_{-2.7} \times 10^{22}\,$cm$^{-2}$, a photon index $\Gamma=1.3^{+0.5}_{-0.4}$ 
and an unabsorbed energy flux $\Phi$(2-10 keV)$= 6.9^{+1.7}_{-2.8} \times 10^{-13}\,$erg$\,$cm$^{-2}\,$s$^{-1}$
with a p-value of 0.75. 
A search for pulsations was performed at the position of XMMU J183245$-$0921539, but no pulsations were found in the data.

A comparison of the absorption along the line of sight obtained from the X-ray spectral model with the 
column depth derived from the  atomic (HI) and molecular ($^{12}$CO, J=1$\rightarrow$0 transition line) gas 
can be used to provide a lower limit on the distance to XMMU J183245$-$0921539, assuming that all absorbing material is 
at the near distance allowed by the Galactic rotation curve.
The Galactic rotation curve model of \cite{hou09} is used to translate the measured velocities into distances.
A lower limit of about 5 kpc on the distance to XMMU J183245$-$0921539 is thus derived.

\vspace{0.5cm}

\subsection{IR counterparts to XMMU J183245$-$0921539}

The 2MASS catalog\footnote{http://www.ipac.caltech.edu/2mass/releases/allsky/} 
shows one infrared source around the position of XMMU J183245$-$0921539 within the systematic pointing error of XMM-\textit{Newton} ($\sim 2''$).
This source, 2MASS J18324516$-$0921545, lies 1.9$''$ away from the center of gravity of 
the X-ray source. No optical counterpart is found, likely due to strong extinction in the galactic plane.
The apparent magnitudes observed in the J, H, K bands are $m_J=15.52 \pm 0.06$, $m_H=13.26 \pm 0.04$ and $m_K=12.17 \pm 0.02$, respectively.
The IR emission could originate from a massive companion to the X-ray compact source, thus forming a binary system, as discussed 
in section \ref{binary}.

\vspace{0.5cm}

\subsection{$^ {13}$CO observations}\label{13CO}

The Galactic Ring Survey (GRS) performed with the Boston University FCRAO telescopes \cite{grs} provides measurements 
of the $^{13}$CO (J=1$\rightarrow$0) transition line covering the velocity range from -5 to 135 km s$^{-1}$ in this 
region. The detection of this line is evidence for the presence of dense molecular clouds that are known to be targets for 
cosmic-rays and hence gamma-ray emitters via neutral pion production and decay or bremsstrahlung emission.
A 0.2$^\circ$-side square region is defined around the source HESS J1832$-$093 to look for molecular 
clouds traced by the $^{13}$CO transition line. Several molecular clouds measured at different radial velocities are 
found in this region. 
The two structures that show the best spatial coincidence with the TeV excess are selected. 
Their velocity ranges are 26 to 31 $\rm km\,s^{-1}$ and 73 to 81 $\rm km\,s^{-1}$ respectively. 
The antenna temperature of each molecular cloud is integrated on the corresponding range. 
The same Galactic rotation curve model as used in section \ref{xmm} \cite{hou09} is assumed to translate the measured velocities into distances,
each velocity corresponding to two possible distances given our position in the Milky Way. 
However, in case of an association with the SNR, only the near distance
(2.3 kpc and 4.5 kpc respectively) 
would be compatible with the distance estimates to the remnant.
Following the approach described in \cite{simon}, the integrated antenna temperatures are used to derive the gas mass of each structure
and the corresponding gas densities (20 cm$^{-3}$ and 62 cm$^{-3}$ respectively).

\section{Discussion}\label{disc}

\subsection{PWN scenario}\label{pwn}

Dedicated XMM-\textit{Newton} observations at the position of HESS J1832$-$093 have revealed the presence of the point-like source XMMU J183245$-$0921539 showing a hard spectrum ($\Gamma=1.3^{+0.5}_{-0.4}$).
The source position in the galactic plane points towards a location inside our galaxy, at a distance $\ge$5 kpc as deduced from the X-ray power-law model.
XMMU J183245$-$0921539 is a serious counterpart candidate for HESS J1832$-$093 because of its proximity with the best-fit position of the TeV emission and its hard spectrum.
A likely scenario would be that both the X-ray and TeV sources stem from a pulsar wind nebula (PWN) powered by a yet unknown pulsar.
Even if the non-thermal aspect of the X-ray emission is not well determined, its hard spectral index is
indicative of an emission from the vicinity of a pulsar (magnetospheric, striped wind \cite{Petri2007},$\,$...). 
Therefore, despite the lack of observed pulsations in the object, we will consider a pulsar origin for XMMU J183245$-$0921539 in the following.
It can be tested whether energetically a PWN scenario plausibly matches with the population of known TeV-emitting PWNe, under the hypothesis that the X-ray emission comes from the pulsar's magnetosphere.

The unabsorbed flux $\Phi_X$(2-10 keV) of the point source for the 
power-law model is used to compute the corresponding luminosity in the same energy band for a distance of 5 kpc: 
$L_X(2-10\, \rm keV) \simeq 2 \times 10^{33}erg\,s^{-1}$.

It can then be translated to an estimate of the \.{E} of the hypothetical pulsar using the relation provided by 
\cite{li2008}.
The estimated spin-down luminosity is about $1.5 \times 10^{37}\,\rm erg\, s^{-1}$ for the same distance.

If we now compute the \.{E}/d$^2$ we obtain a value of $6 \times 10^{35} \rm erg s^{-1} kpc^{-2}$, 
corresponding to the band for which 70$\%$ of the PWNe are detected by H.E.S.S. \cite{Carrigan}. 
Therefore, if the putative pulsar powers a TeV PWN, it should be detectable by the H.E.S.S. array.
Together with the absence of detected X-ray pulsations, the PWN scenario remains unconfirmed but is energetically possible.

\vspace{0.3cm}

\subsection{Binary scenario}\label{binary}

The infrared source 2MASS J18324516$-$0921545 discovered in spatial coindidence with XMMU J183245$-$0921539
could suggest that the X-ray source resides in a binary system around a massive star.
The magnitude extinction number in the optical band expected from the X-ray column density is A$_{\rm V}$=59$^{+17}_{-15}$, using the $N_H-A_{\rm V}$ relation given by \cite{Predehl1995}.
This would result in an absolute magnitude M$_{\rm J} \le$-14.6 for a distance $\ge$5 kpc \cite{Cardelli1989}. This value is excluded but the $N_H-A_{\rm V}$ relation cannot apply in case of strong local absorption.
Therefore, a binary scenario with 2MASS J18324516$-$0921545 as optical companion could only work if the X-ray absorption arises mainly locally around XMMU J183245$-$0921539.
In the absence of orbitally modulated X-ray or TeV emission, the binary possibility remains unconfirmed for the moment. 

\vspace{0.3cm}

\subsection{SNR-molecular cloud interaction scenario}

Despite the lack of X-ray emission from the SNR shell that could be due to synchrotron radiation of high energy electrons,
or thermal X-ray emission which could stem from shock-gas interactions 
seen frequently in middle-aged SNRs, the observed VHE emission might still come from particles accelerated in the remnant.
Those particles would then interact with localized target material via neutral pion production and decay or bremsstrahlung emission.
Indeed, $^{13}$CO measurements show the presence of structures around HESS J1832$-$093, while at corresponding distance ranges lower gas densities are seen in other portions of the SNR shell.
There is, however, no support of the association of these gas structures with G22.7$-$0.2, e.g. through maser emission from shock-cloud interaction. 
An identification with the TeV source might therefore be due to chance coincidence.
Further investigation on a possible CR propagation to a nearby MC at the origin of the gamma-ray emission thus needs to be performed.

\section{Conclusion}\label{conclusions}

Observations in the field of view of SNR G22.7$-$0.2 have led to the discovery of the VHE source HESS J1832$-$093 lying on the edge of the 
SNR radio rim. 
The available multi-wavelength data, using archival radio and infrared data, as well as a dedicated XMM-\textit{Newton} X-ray pointing towards the source,
do not permit to unambiguously determine the nature of the object that gives rise to the VHE emission.
A compelling X-ray counterpart, XMMU J183245$-$0921539, has been discovered. The nature of this X-ray source could, however, not be established from the X-ray data alone.
Together with the TeV emission and the infrared point source  2MASS J18324516$-$0921545, plausible object classifications are a pulsar wind nebula or a binary system.
Cosmic-rays accelerated in the SNR G22.7$-$0.2 interacting with dense gas material could also result in TeV emission but the interaction between the SNR and a molecular cloud
is not supported by observational evidence such as maser emission.

The nature of the gamma-ray source HESS J1832$-$093 presented here remains, therefore, undetermined.
More data in X-rays and radio to look for variability and faint diffuse emission could help to constrain the nature of the source.

\section*{Acknowledgments}

\small{The support of the Namibian authorities and of the University of Namibia
in facilitating the construction and operation of H.E.S.S. is gratefully
acknowledged, as is the support by the German Ministry for Education and
Research (BMBF), the Max Planck Society, the German Research Foundation (DFG), 
the French Ministry for Research,
the CNRS-IN2P3 and the Astroparticle Interdisciplinary Programme of the
CNRS, the U.K. Science and Technology Facilities Council (STFC),
the IPNP of the Charles University, the Czech Science Foundation, the Polish 
Ministry of Science and  Higher Education, the South African Department of
Science and Technology and National Research Foundation, and by the
University of Namibia. We appreciate the excellent work of the technical
support staff in Berlin, Durham, Hamburg, Heidelberg, Palaiseau, Paris,
Saclay, and in Namibia in the construction and operation of the
equipment.}


\begin{thebibliography}{}

\bibitem{crab} H.E.S.S. collaboration, AAP 457 (2006) 899-915 doi:10.1051/0004-6361:20065351
\bibitem{Shaver} Shaver, P.~A. and Goss, W.~M., AJPAS 14 (1970) p.133 Bib. code:1970AuJPA..14..133S
\bibitem{parismva} Becherini, Y. et al., Astro-Ph. 34 (2011) 858-870 doi:10.1016/j.astropartphys.2011.03.005
\bibitem{hess2006survey}  H.E.S.S. collaboration, ApJ 636 (2006) 777-797 doi:10.1086/498013
\bibitem{magpis} Helfand, D.~J. et al., AJ 131 (2006) 2525-2537 doi:10.1086/503253
\bibitem{carter07} Carter, J.~A. and Read, A.~M., AAP 464 (2007) 1155-1166 doi:10.1051/0004-6361:20065882
\bibitem{hou09} Hou, L.~G. and Han, J.~L. and Shi, W.~B., AAP 499 (2009) 473-482 doi:10.1051/0004-6361/200809692
\bibitem{grs} Jackson, J.~M. et al., ApJS 163 (2006) 145-159 doi:10.1086/500091
\bibitem{simon} Simon, R. et al., ApJ 551 (2001) 747-763 doi:10.1086/320230
\bibitem{Petri2007} P{\'e}tri, J. and Lyubarsky, Y., AAP 473 (2007) 683-700 doi:10.1051/0004-6361:20066981
\bibitem{li2008} Li, X.-H. and Lu, F.-J. and Li, Z., ApJ 682 (2008) 1166-1176 doi:10.1086/589495
\bibitem{Carrigan} Carrigan, S. et al. for the H.E.S.S. collaboration, proc. ICRC (2007) eprint:arXiv:0709.4094v1
\bibitem{Predehl1995} Predehl, P. and Schmitt, J.~H.~M.~M., A$\&$A 293 (1995) 889-905 Bib. code:1995A$\&$A...293..889P
\bibitem{Cardelli1989} Cardelli, J.~A. and Clayton, G.~C. and Mathis, J.~S., proc. IAU (1989) Bib. code:1989IAUS..135P...5C
\bibitem{dav} Drury, L.~O. and Aharonian, F.~A. and Voelk, H.~J., AAP 287 (1994) 959-971 Bib. code:1994A$\&$A...287..959D

\end{thebibliography}
\end{document}